\documentclass[floats,floatfix,showpacs,amssymb,prd,twocolumn,superscriptaddress,nofootinbib]{revtex4-1}

\usepackage{graphicx,epsf, epsfig, amssymb}
\usepackage{bm}
\usepackage{longtable}
\usepackage[usenames,dvipsnames]{xcolor}
\usepackage{color}
\usepackage{hyperref}
\usepackage{amsfonts,amsmath,amssymb,mathrsfs}
\usepackage{multirow}
\usepackage[nolist,nohyperlinks]{acronym}
\usepackage{xspace}
\usepackage{booktabs,array}
\usepackage[british]{babel}
\usepackage[utf8]{inputenc}

\def\be{\begin{equation}}
\def\ee{\end{equation}}
\def\beq{\begin{eqnarray}}
\def\eeq{\end{eqnarray}}

\begin{document}

\title{Gravity-dominated unequal-mass black hole collisions}

\author{Ulrich Sperhake}
\email{U.Sperhake@damtp.cam.ac.uk}
\affiliation{Department of Applied Mathematics and Theoretical Physics,
Centre for Mathematical Sciences, University of Cambridge,
Wilberforce Road, Cambridge CB3 0WA, UK}
\affiliation{California Institute of Technology, Pasadena, CA 91109, USA}
\affiliation{Department of Physics and Astronomy, The University of
Mississippi, University, MS 38677, USA}
\author{Emanuele Berti}
\email{eberti@olemiss.edu}
\affiliation{Department of Physics and Astronomy, The University of
Mississippi, University, MS 38677, USA}
\affiliation{CENTRA, Departamento de F\'{\i}sica, Instituto Superior
T\'ecnico, Universidade de Lisboa - UL, Av.~Rovisco Pais
1, 1049 Lisboa, Portugal}
\author{Vitor Cardoso}
\email{vitor.cardoso@ist.utl.pt}
\affiliation{CENTRA, Departamento de F\'{\i}sica, Instituto Superior
T\'ecnico, Universidade de Lisboa - UL, Av.~Rovisco Pais
1, 1049 Lisboa, Portugal}
\affiliation{Perimeter Institute for Theoretical Physics, 31 Caroline Street North,
Waterloo, Ontario N2L 2Y5, Canada}
\affiliation{Dipartimento di Fisica, ``Sapienza'' Universit\`a di Roma \& Sezione INFN Roma1, P.A. Moro 5, 00185, Roma, Italy}
\affiliation{Department of Physics and Astronomy, The University of
Mississippi, University, MS 38677, USA}
\author{Frans Pretorius}
\email{fpretori@princeton.edu}
\affiliation{Department of Physics, Princeton University, Princeton,
NJ 08544, USA}

\pacs{04.25.dg, 04.70.Bw, 04.30.-w}

\date{\today}

\begin{abstract}
  We continue our series of studies of high-energy collisions of black
  holes investigating unequal-mass, boosted head-on collisions in four
  dimensions. We show that the fraction of the center-of-mass energy
  radiated as gravitational waves becomes independent of mass ratio
  and approximately equal to $13\%$ at large energies. We support this
  conclusion with calculations using black hole perturbation theory
  and Smarr's zero-frequency limit approximation.  These results lend
  strong support to the conjecture that the detailed structure of the
  colliding objects is irrelevant at high energies.
\end{abstract}

\maketitle

\noindent
{\bf \em Introduction.}
Numerical simulations of black hole (BH) collisions are an ideal
framework to understand the behavior of gravity in the strong-field
regime. These simulations allow us to answer fundamental questions and
to verify (or disprove) some of our cherished beliefs about Einstein's
general relativity (GR). Are BH collisions subject to cosmic
censorship, so that naked singularities are never the outcome of any
such event?  What is the upper limit of the fraction of kinetic energy
of the system that can be radiated in gravitational waves (GWs) during
these collisions? In the ultrarelativistic (UR) limit, what properties
of the collision, if any, are dependent on the underlying structure of
the colliding objects, here the spins of the BHs and their mass ratio?

Some years ago we started a long-term program to answer these
questions. We first showed that the head-on collision of two
equal-mass BHs at the speed of light will radiate no more than
$\sim 14\pm 3\%$ of the energy of the system~\cite{Sperhake:2008ga}
(this result was recently confirmed independently by the RIT
group~\cite{Healy:2015mla}, refining the limit to $13 \pm 1\%$).  This
is less than half the upper limit of $\sim 29\%$ predicted by Penrose
in the seventies, but two orders of magnitude larger than the energy
radiated when two BHs collide head-on from rest
\cite{Sperhake:2005uf}.  We found that collisions with finite impact
parameter can be tuned to exhibit ``zoom-whirl''
behavior~\cite{Glampedakis:2002ya,Pretorius:2007jn} and that they can
produce near-maximally spinning remnants~\cite{Sperhake:2009jz}.  We
also used zero-frequency limit (ZFL) calculations pioneered by
Smarr~\cite{Smarr:1977fy} and BH perturbation theory to clarify the
structure of the radiation~\cite{Berti:2010ce}.
We studied grazing collisions with aligned spins, showing that in the
UR limit the radiated energy and scattering thresholds become
spin-independent~\cite{Sperhake:2012me}.  In principle, in the UR
limit it may be possible to radiate all kinetic energy as GWs by
fine-tuning the collision near threshold, but extrapolations of our
numerical results suggest that there is an upper limit of $\sim$50\%
on the radiation that can be emitted (this number is consistent with
perturbative calculations in the extreme-mass-ratio limit presented
in~\cite{Gundlach:2012aj}). By analyzing the evolution of the apparent
horizon, we found that the other half of the kinetic energy is
absorbed and converted into rest mass of the merger remnant, or of the
scattering constituents in non-merging cases. Furthermore, we
demonstrated that non-merging scattering events with spinning BHs can
exhibit large center-of-mass recoil velocities due to GW emission,
showing that the formation of a common horizon is not necessary to
impart a kick to the system~\cite{Sperhake:2010uv}.  All of our
calculations support Penrose’s cosmic censorship conjecture. Note
however that the quoted results have been from studies in four
spacetime dimensions ($D=4$); Ref.~\cite{Okawa:2011fv} presented
evidence {\em suggesting} that high-speed collisions in $D=5$ can
lead to naked singularities from a generic subset of initial
conditions.

One of the main conclusions to be drawn from our simulations of
equal-mass, spinning BH collisions is that spin does not matter in the
high-energy limit. In GR and in $D=4$, isolated BHs in vacuum are
uniquely characterized by their masses and spins. Since classical GR
has no intrinsic scale then, beyond spin the only way this
``ultraviolet universality'' may be violated is by varying the mass
ratio.  Here we investigate one aspect of this problem by asking the
following question: does the binary mass ratio affect the maximum
amount of energy that can be radiated in UR head-on collisions? This
paper bridges the gap between our previous simulations of UR,
equal-mass head-on collisions~\cite{Sperhake:2008ga} and the
simulations of~\cite{Sperhake:2011ik}, which considered
nonrelativistic head-on collisions with mass ratios as small as
$q=1/100$. The main product of this work is another confirmation of
the simplicity and elegance of UR collisions in GR: we find that the
maximum fraction of the total energy radiated as GWs in this limit is
$\sim 0.13$, irrespective of the binary mass ratio.

A consequence of this research, along with previous studies of
high-energy collisions of
``stars''~\cite{Choptuik:2009ww,East:2012mb,Rezzolla:2012nr}, is more
solid evidence that the structure of the colliding objects is
irrelevant at large energies. In other words, in this regime the
outcome of colliding objects with a complex multipolar structure is
equivalent to colliding Schwarzschild BHs. Any intricacies associated
with matter interactions are hidden behind horizons and do not leave a
distinguishable imprint on the GW signal.

\smallskip
\noindent
{\bf \em Setup}:
Consider the collision of two nonspinning, electrically neutral BHs with rest
masses $m_{A,B}$, total rest mass $M_0 \equiv m_A+m_B$, and mass ratio
$q \equiv m_A / m_B \le 1$.  In the center of mass (CM) frame,
where we measure all quantities,
the velocity of BH $A$ is $v_{A}$
with corresponding Lorentz factor
$\gamma_{A} = (1-v_{A}^2)^{-1/2}$, and we can define its
energy and momentum as $E_A = \gamma_A m_A$ and $P_A = \gamma_A m_A v_A$
respectively; likewise for BH $B$. In terms of these quantities, the CM
frame is defined by
\begin{equation}
   P \equiv P_A + P_B = m_B(q\gamma_A v_A + \gamma_B v_B) = 0\,.
\end{equation}
The total energy of the spacetime is defined as $M\equiv E_A+E_B$, and
we further introduce an effective Lorentz factor $\gamma$ such that
\begin{equation}
\gamma M_0 \equiv M = E_A + E_B = m_B(q\gamma_A + \gamma_B)\,.
\end{equation}
In other words, $1-1/\gamma$ measures the fraction of total energy that is initially
in the form of kinetic energy.

In terms of parameters characterizing the initial data of each
similation, those most relevant to the collision problem are (i) the
mass ratio $q$, which in this study takes the values
$q=1,\,1/2,\,1/4,\,1/10$, and (ii) the effective Lorentz factor
$\gamma$, or equivalently the velocities $v_A$ and $v_B$. We use
initial separations in the range $d/M\approx 40$ to $d/M\approx 100$,
observing that for such large values the radiation is essentially
independent of $d$.  The key diagnostic quantity is the amount of
energy $E_{\rm rad}$ radiated in GWs, normalized to the total
spacetime mass $M$, and excluding a contribution from an early burst
of spurious (``junk'') radiation coming from the initial data.

Equal-mass collisions with $q=1$ were discussed
in~\cite{Sperhake:2008ga}. Here we perform additional simulations of
BH collisions with unequal masses using the {\sc Lean} code
\cite{Sperhake:2006cy} which is based on {\sc Cactus}
\cite{Cactusweb,Allen:1999} und uses {\sc Carpet}
\cite{Carpetweb,Schnetter:2003rb} for mesh refinement, {\sc
  AHFinderDirect} \cite{Thornburg:1995cp,Thornburg:2003sf} and the
spectral solver of Ref.~\cite{Ansorg:2004ds} for initial data
generation.  For these new simulations, we fix the resolution by the
scale $m_A$ of the smaller hole to $h=m_A/80$ near the BH
singularities, and increase it by a factor 2 on each consecutive outer
refinement level, for a total of 10 refinement levels when
$q=1/2,~1/4$, or 12 levels when $q=1/10$.  To measure gravitational
radiation we compute the Newman-Penrose scalar $\Psi_4$ at several
radii, typically within a range $[50\ldots 200]M$.  We then decompose
$\Psi_4$ into multipole modes $\psi_{lm}$ of the spherical harmonics
${_{-2}}Y_{lm}$ of spin-weight $-2$:
\be
\Psi_4(t,r,\theta,\phi)=\sum_{l=2}^\infty \sum_{m=-l}^l
\,{_{-2}}Y_{lm}(\theta\,,\phi)\, \psi_{lm}(t,r)\,.
\ee
Due to the symmetries of this problem, the only nonvanishing
multipoles all have $m=0$. The energy flux is then given by
\be 
\dot{E}=\sum_l
\lim_{r\rightarrow \infty}\frac{r^2}{16\pi} \left|
\int_{-\infty}^{t} \psi_{l0}(\tilde{t}) d\tilde{t} \right|^2
\equiv \sum_l \dot{E_l}\,,
\label{luminosity}
\ee
where overdots ($\dot{\ }$) denote time derivatives.

Our results are affected by three main sources of uncertainty: the
finite extraction radius $r_{\rm ex}$, the discretization error and
the spurious initial radiation. We estimate the error arising from
using a finite extraction radius by measuring the waveform components
at several radii, and fitting the resultant flux to an expression of
the form
$\dot{E}(r,t)=\dot{E}^{(0)}(t)+\dot{E}^{(1)}(t)/r$.
The estimated uncertainty is then given by the difference between the
net radiated energy $E_{\rm rad}$ calculated using the extrapolated result 
$\dot{E}^{(0)}$ and that calculated with $\dot{E}(r_{\rm ex})$ at the largest value of
$r_{\rm ex}$. We find that the fractional uncertainty in $E_{\rm rad}$
amounts to a maximum of $2\%$ for
low or vanishing boosts, $4\%$ for moderate velocities
$v_A \approx 0.5$, and $8\%$ for the largest boosts $v_A\approx 0.9$.

\begin{figure}[bht]
  \centering
  \includegraphics[width=\columnwidth,clip=true]{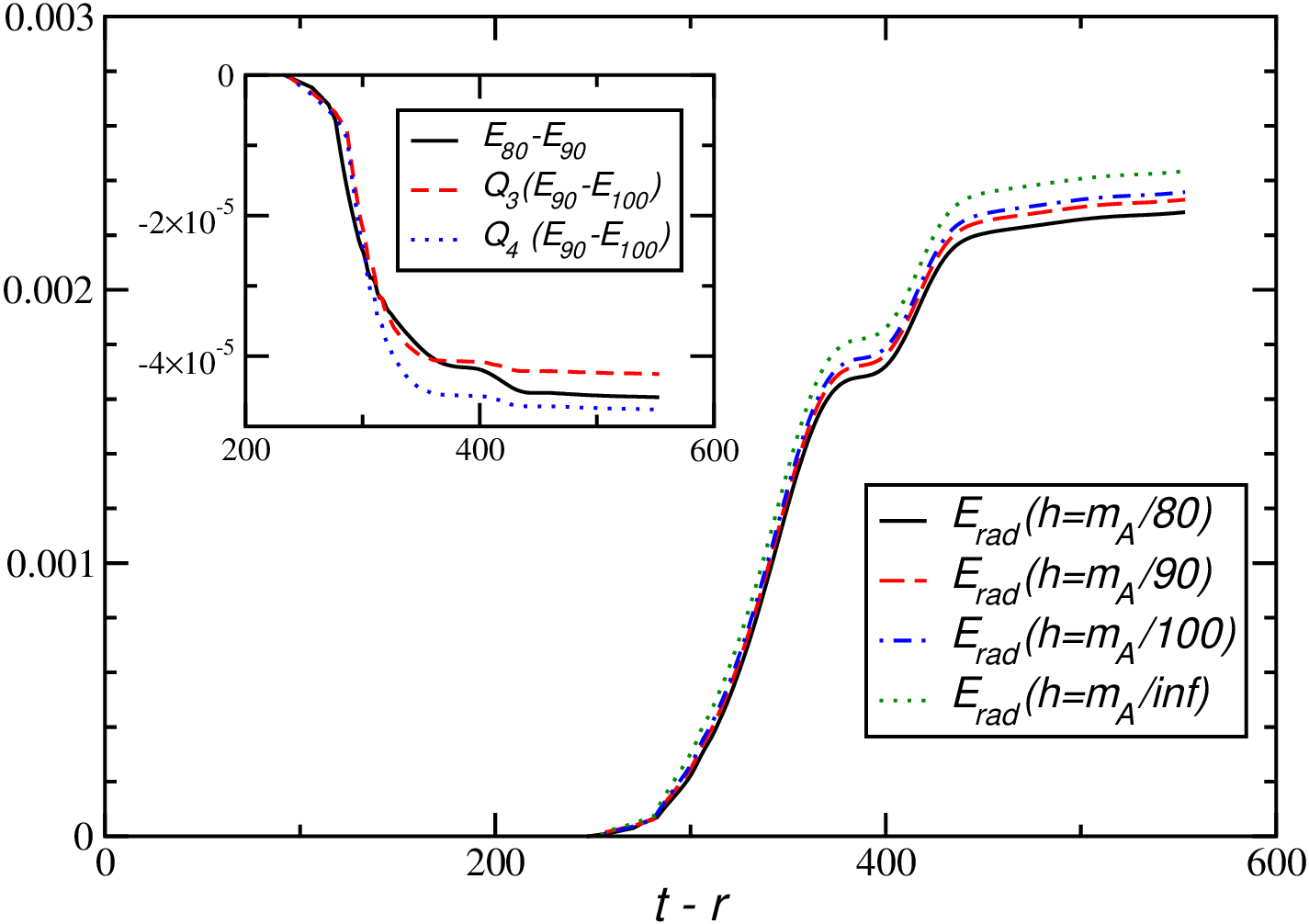}
  \caption{Convergence plot for a $q=0.10$, $\gamma=1.11$
    binary where $v_A=0.87$. The radiated energy is shown in units
    of the total mass $M$. The inset shows the deviations between
    coarse, medium and high resolution rescaled for third-order
    convergence ($Q_3=1.56$) and fourth-order convergence
    ($Q_4= 1.75$). The dotted curve in the main panel shows the
    energy extrapolated to infinite resolution using the more
    conservative 3rd-order estimate.}
  \label{fig:convergence}
\end{figure}

To estimate discretization errors we have evolved one of the most
challenging collisions, namely $q=1/10$ and $v_A=0.87$, using
additional resolutions $h=m_A/90$, $m_A/100$ (with all coarser
refinement levels adjusted accordingly). Figure~\ref{fig:convergence}
shows that the quantity $E_{\rm rad}$ exhibits between third- and
fourth-order convergence.  A conservative estimate obtained assuming
third-order convergence gives a fractional error of $6\%$, and we
complement this with estimates of $2\%$ obtained from the prior
$\gamma=1$ study in~\cite{Sperhake:2011ik}.

Finally, the conformally flat puncture initial data contain spurious
gravitational radiation, which increases strongly with boost $\gamma$.
In order to extract physically meaningful information we must separate
the spurious radiation from the radiation generated by the collision
itself. This is done by ``waiting'' for the spurious radiation to pass
the last extraction radius, and then discarding the earlier part of
the GW signal.  The exact choice of the time where to separate
spurious initial radiation from that generated in the collision itself
introduces an uncertainty, which we estimate by varying this choice
guided by the quadrupole radiation, where spurious and physical
radiation can be identified most clearly. For low boosts the
resulting error is negligible, but it increases significantly to $6\%$
for $v_A\approx 0.6$ and $10\%$ $(12\%)$ for $q=1/2$ $(q=1/4,~1/10)$
at $v\approx 0.9$.  In summary, the total error budget is about $4\%$
when $v_A=0$, $14\%$ when $v\approx 0.6$ and $24\%~(26\%)$ when
$v_A\approx 0.9$ for $q=1/2~(1/4,~1/10)$.

\begin{figure}[thb]
  \includegraphics[width=\columnwidth,clip=true]{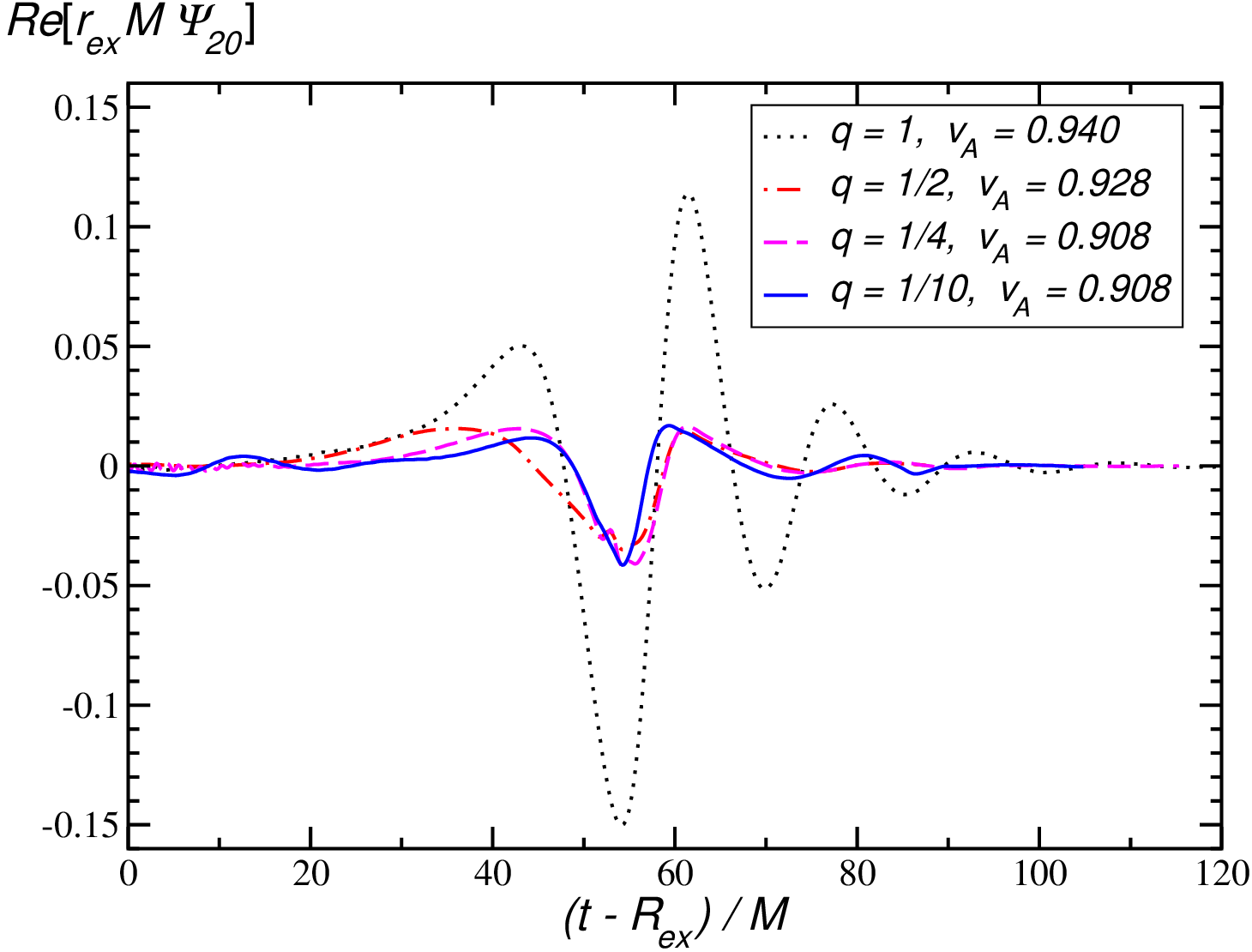}
  \caption{The dominant multipole $\psi_{20}$ of the Newman-Penrose
    scalar extracted from the most relativistic collisions considered
    for each mass ratio. The imaginary part of $\psi_{20}$ vanishes
    for all cases due to symmetry.}
  \label{fig:flux}
\end{figure}

\smallskip
\noindent
{\bf \em Results.}
The waveforms and corresponding energy fluxes from a set of the most
challenging runs are shown in Fig.~\ref{fig:flux}. The waveforms have 
a structure familiar in BH dynamics~\cite{Sperhake:2008ga}:
a precursor, a main burst at the onset of the formation of a common apparent horizon,
and a final ringdown tail. We find that, to a good approximation, the final BH rings down 
in the lowest QNM frequency as predicted by linear theory~\cite{Berti:2005ys,Berti:2009kk}.

\begin{figure}[ht]
  \centering
  \includegraphics[width=\columnwidth,clip=true]{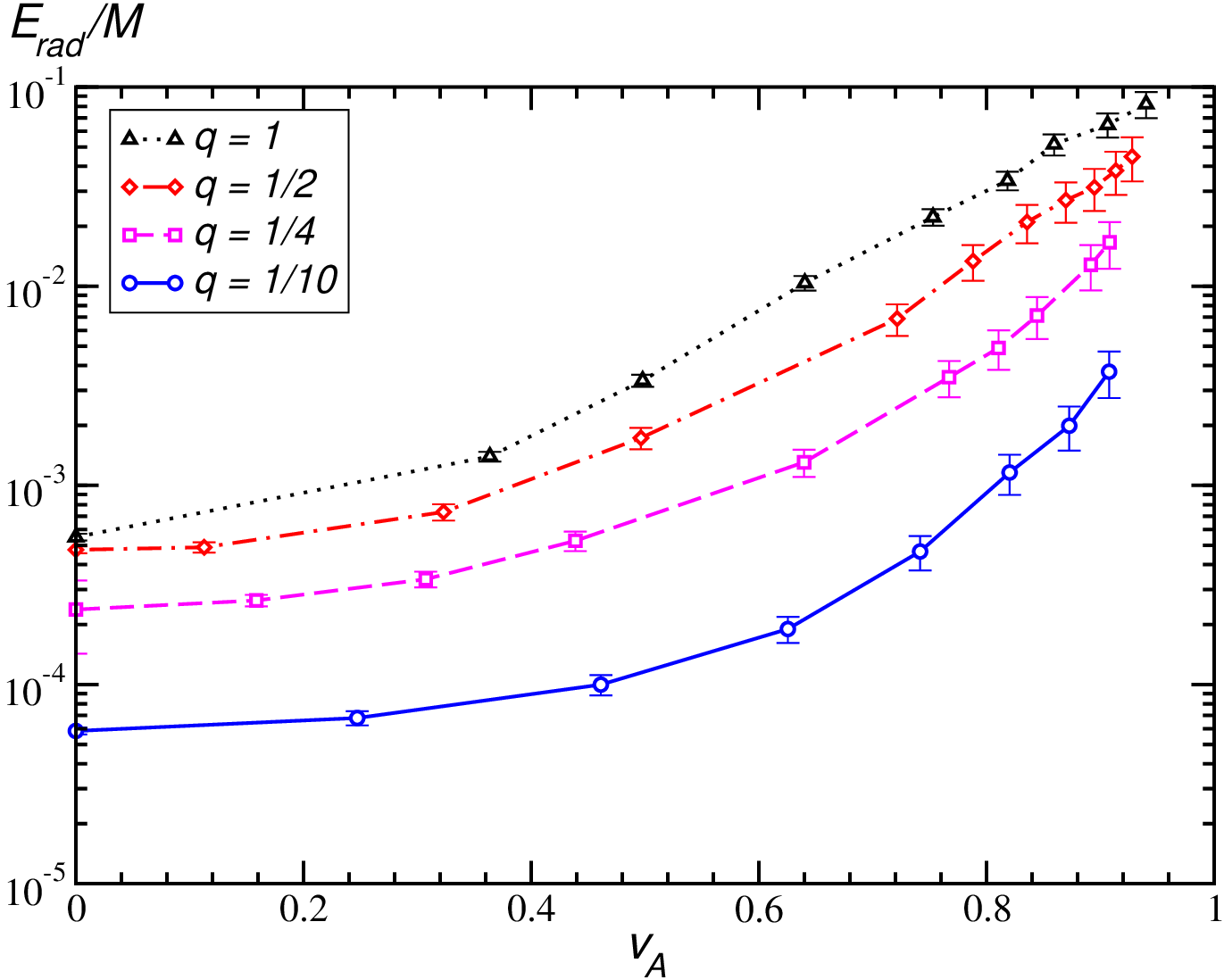}
  \caption{Total energy radiated $E_{\rm rad}/M$ as a function of 
    $v_A$.}
  \label{fig:erad}
\end{figure}

The total integrated energy $E_{\rm rad}$ radiated in GWs
(normalized by the total center-of-mass energy $M$) is shown in
Fig.~\ref{fig:erad} for all the simulations we studied.
This quantity rapidly increases for large boosts.  To understand the
limiting behavior we resort to two perturbative calculations: the ZFL
and point-particle approximations.

The ZFL~\cite{Berti:2010ce} has
been very successful at describing the functional dependence of the
nonlinear results for equal-mass collisions~\cite{Sperhake:2008ga,Healy:2015mla}. 
We therefore use the ZFL result for generic {\em unequal-mass}
collisions~\cite{Smarr:1977fy}, in particular the spectrum per unit
solid angle
\beq
 f(\theta,q,v_A)&\equiv& \frac{1}{\gamma_A^2m_A^2}\frac{d^2E_{\rm
     rad}}{d\omega d\Omega} \\
&=&\frac{v_A^2\sin^4\theta}{4\pi^2}\left[\frac{v_A+v_B}{(1-v_A\cos\theta)(1+v_B\cos\theta)}\right]^2\,.
\nonumber
\eeq
With the physically reasonable assumption that there is a cutoff
frequency at $\omega_c\sim X(q)/M$, we get
\be
\frac{E_{\rm rad}}{M}=X(q)\frac{m_A^2\gamma_A^2}{M^2} F(v_A,q)\,,\label{energy_zfl}
\ee
where $F(v_A,q)=\int d\Omega f(\theta,q,v_A)$ can be computed
analytically. In other words, the ZFL gives an analytical prediction
with only one unknown parameter $X(q)$, and for very large CM energies
$E_{\rm rad}/M\to X(q)/\pi$. By fitting the last three points in
Fig.~\ref{fig:erad} for each value of $q$ to Eq.~\eqref{energy_zfl} we
can get the percentage of energy radiated for each mass ratio in the
UR limit, $\epsilon(q)\equiv 100 X(q)/\pi$:
\beq
&\epsilon(1)=12.7\pm 1.5\,, \quad
\epsilon(1/2)=11.2 \pm 2.7\,, \quad \nonumber\\
&\epsilon(1/4)=11.6 \pm 3.0\,, \quad
\epsilon(1/10)=12.0\pm 3.0\,.
\eeq
Our results for $\epsilon(q)$ consistently lie in the $11-13\%$
interval for all mass ratios.  This strongly supports the conjecture
that the structure of the colliding objects becomes irrelevant at
large center-of-mass energies.

There is another limit which is amenable to a semi-analytic treatment,
and that is when a small BH $A$ with energy $E_A$ collides with a
large BH with mass $M$ such that $E_A \ll M$. This is the
point-particle
limit~\cite{Davis:1971gg,Cardoso:2002ay,Berti:2003si,Berti:2010ce,Berti:2010gx}. Nonlinear
head-on collision results agree extremely well with point-particle
predictions even when the mass ratios in the simulations approach
unity, at least for low-energy encounters~\cite{Sperhake:2011ik}. To
test whether they also agree for high-energy collisions we consider,
as a representative example, our smallest mass-ratio runs with
$q=1/10$ (which are marginally within the regime of validity of
perturbation theory). By computing the radiation for a point-like
particle with velocity $v_A=0.91$ and energy $E_A=2.38m_0$ (where
$m_0$ is the particle's rest mass) falling into a massive BH of mass
$M$ through a numerical integration of the Zerilli equation for
multipoles up to $l=6$ with two independent codes we find
\be\label{EnPP}
\frac{M E_{\rm rad}}{E_A^2}=0.090\,.
\ee
To make contact with our results, we take $M$ to be the CM energy
$M=E_A+E_B$ (we could equally well take $M=E_B$, as the two are
equivalent in the point-particle limit; this intrinsic ambiguity would
not affect the agreement between the point-particle results and the
numerics shown below).
Our full nonlinear, numerical simulations for $q=1/10$ yield
$(E_B+E_A) E_{\rm rad}/E_A^2=0.104$.  This surprisingly good agreement
($\sim 10\%$) provides further support to our results.

\smallskip
\noindent
{\bf \em Conclusions.}
The main conclusion of the present study is to confirm the
expectation, borne out of our previous
work~\cite{Sperhake:2008ga,Sperhake:2009jz,East:2012mb,Sperhake:2012me},
that the structure of the colliding objects does not matter in
gravity-dominated collisions. We have previously demonstrated that the
effect of spin on the radiated energy and scattering threshold becomes
negligible for grazing collisions in the UR limit
\cite{Sperhake:2012me}. In this work we show that the effect of the
binary mass ratio on the radiated energy becomes negligible in the UR
limit for head-on collisions. It will be interesting to verify whether
the effect of mass ratio is likewise irrelevant in the UR limit for
{\em grazing} collisions.

Another interesting extension will be to consider spacetime dimensions
$D>4$. Following the first simulations in $D=5$~\cite{Witek:2010az}
and $D=6$~\cite{Witek:2014mha}, progress on UR collisions in higher
dimensions has been slower than expected, due to technical
complications in achieving code stability. A new formulation of the
higher-dimensional Einstein equations now allows our {\sc Lean} code
to compute the radiation from head-on collisions up to $D=10$, so that
the number of extra dimensions considered possible in
higher-dimensional gravity scenarios~\cite{Kanti:2004nr} falls within
the range achievable by the code.
The study of unequal-mass UR collisions in higher dimensions is of
particular interest because perturbative calculations suggest that the
percentage of kinetic energy radiated in GWs may reach a minimum as a
function of $D$, and then increase again~\cite{Berti:2010gx}. The
perturbative calculations do not hold for $D\geq13$, since they
predict a total radiation output which breaks the assumptions behind
the formalism~\cite{Berti:2010gx}, and therefore our understanding of
radiation in large-$D$ spacetimes is still lacking. We plan to
investigate this problem in the near future.

\smallskip
\noindent
{\bf \em Acknowledgments.}
This work was supported by the H2020-MSCA-RISE-2015 Grant No. StronGrHEP-690904,
the SDSC Comet and TACC Stampede clusters through
NSF-XSEDE Grant No.~PHY-090003,
STFC Consolidator Grant No. ST/L000636/1,
and
DiRAC's Cosmos Shared Memory system through BIS Grant No.~ST/J005673/1
and STFC Grant Nos.~ST/H008586/1, ST/K00333X/1.
E.B. is supported by NSF CAREER Grant No. PHY-1055103 and by FCT
contract IF/00797/2014/CP1214/CT0012 under the IF2014 Programme.
V.C. thanks the Departament de F\'{\i}sica Fonamental at Universitat de Barcelona for hospitality while this work was being completed.
V.C. and U.S. acknowledge financial support provided under the European
Union's H2020 ERC Consolidator Grant ``Matter and strong-field
gravity: New frontiers in Einstein's theory'' grant agreement
no. MaGRaTh--646597.
V.C. also acknowledges financial support from FCT under Sabbatical
Fellowship nr. SFRH/BSAB/105955/2014.
F.P. acknowledges financial support from the Simons
Foundation and NSF grant PHY-1305682.
This research was supported in part by the Perimeter Institute for
Theoretical Physics. Research at Perimeter Institute is supported by
the Government of Canada through Industry Canada and by the Province
of Ontario through the Ministry of Economic Development $\&$
Innovation.

\bibliographystyle{apsrev}

\end{document}